\documentclass[12pt]{article}
 \usepackage{epsfig}
 \def\be{\begin{equation}}
 \def\ee{\end{equation}}
 \def\bea{\begin{eqnarray}}
 \def\eea{\end{eqnarray}}
 \usepackage{graphicx}

 \catcode`\@=11
 \def\lsim{\mathrel{\mathpalette\@versim<}}
 \def\gsim{\mathrel{\mathpalette\@versim>}}
 \def\@versim#1#2{\vcenter{\offinterlineskip
 \ialign{$\m@th#1\hfil##\hfil$\crcr#2\crcr\sim\crcr } }}
 \catcode`\@=12

 \catcode`@=12
\begin{document}
 \thispagestyle{empty}
 \begin{flushright}
 UCRHEP-T606\\
 Dec 2020\
 \end{flushright}
 \vspace{0.6in}
 \begin{center}
 {\LARGE \bf One-Loop Electron Mass and\\ 
 Three-Loop Dirac Neutrino Masses\\}
 \vspace{1.2in}
 {\bf Ernest Ma\\}
 \vspace{0.2in}
{\sl Physics and Astronomy Department,\\ 
University of California, Riverside, California 92521, USA\\}
\end{center}
 \vspace{1.2in}

\begin{abstract}
In the context of a left-right extension of the standard model of quarks 
and leptons with the addition of a gauged $U(1)_D$ dark 
symmetry, it is shown how the electron may obtain a radiative mass in one 
loop and two Dirac neutrinos obtain masses in three loops.  
\end{abstract}

\newpage
\baselineskip 24pt
\noindent \underline{\it Introduction}~:~
Neutrino masses are very small.  In the context of the minimal standard 
model (SM) of particle interactions, it has a natural explanation because they 
must come from a dimension-five operator~\cite{w79}, i.e.
\begin{equation}
{\cal L}_5 = {f_{ij} \over 2 \Lambda} (\nu_i \phi^0 - l_i \phi^+) 
(\nu_j \phi^0 - l_j \phi^+), 
\end{equation}
where $(\nu,l)_{1,2,3}$ are the left-handed lepton doublets of the three 
families, and $(\phi^+,\phi^0)=\Phi$ is the one Higgs scalar doublet. 
Neutrino masses are thus Majorana and inversely proportional to the large 
scale $\Lambda$, hence the name ``seesaw''.  There are exactly three 
ways~\cite{m98} to realize ${\cal L}_5$ at tree level, establishing the 
nomenclature Type I,II,III seesaw, as well as three ways to realize it 
radiatively in one loop~\cite{m98}.  One particular application is to 
let dark matter~\cite{bh18} be the origin of radiative neutrino mass, 
as in the ``scotogenic'' model~\cite{m06}. 

If neutrinos are Dirac particles, thereby requiring the existence of $\nu_R$ 
and the maintenance of a conserved lepton number $L$, the SM is not a 
natural accommodating framework.  First, under 
$SU(3)_C \times SU(2)_L \times U(1)_Y$, $\nu_R$ is trivial so its 
existence is not very well justified.  Second, the $L$ symmetry must be 
imposed to forbid the otherwise allowed $\nu_R$ Majorana mass.  Third, 
the Yukawa coupling linking $\nu_L$ to $\nu_R$ through $\phi^0$ must 
be chosen to be of order $10^{-12}$ to account for the data on neutrino 
masses.  To alleviate this last shortcoming, a symmetry is often employed 
to forbid the offending Yukawa term, say $Z_2$ under which $\nu_R$ is odd. 
At the same time, $L$ is still assumed to be exact, wheras $Z_2$ will be 
broken softly by the addition of other fermions and scalars.  This 
procedure~\cite{mp17} has been studied in a variety of models.

Consider now the left-right extension of the SM.  The existence of $\nu_R$ 
is required as part of an $SU(2)_R$ doublet.  Neutrinos obtain Dirac 
masses in the same way as the other fermions through a scalar bidoublet. 
If $SU(2)_{L,R}$ doublets $\Phi_{L,R}$ are also added to break 
$SU(2)_L \times SU(2)_R \times U(1)_{B-L}$ to $U(1)_Q$, then $L$ remains 
a global symmetry.  The first and second shortcomings of the SM for Dirac 
neutrinos no longer apply, but the third remains.  A well-known 
approach~\cite{bms03} is to keep only $\Phi_{L,R}$ without the scalar 
bidoublet.  Hence all fermion masses are zero at this point.  By adding 
heavy singlet fermions, the known quarks and leptons may now obtain 
seesaw masses~\cite{dw87}.  Neutrinos may also be chosen to have radiative 
Dirac masses~\cite{m88,m89,ms18,m20}.  Recently, it has been 
shown~\cite{m20-1} that all fermion masses could be radiative if the 
$SU(2)_R$ breaking scale is high enough.  In this work, a new and different 
scenario is envisioned: the emergence of three-loop neutrino masses 
in the presence of a one-loop electron mass, with the help of a dark 
gauged $U(1)_D$ symmetry.

\noindent \underline{\it Description of Model}~:~
The particle content of the proposed model is listed in Table 1.
\begin{table}[tbh]
\centering
\begin{tabular}{|c|c|c|c|c|c|}
\hline
fermion/scalar & $SU(3)_C$ & $SU(2)_L$ & $SU(2)_R$ & $U(1)_{B-L}$ & $U(1)_D$ \\
\hline
$(\nu_e,e)_L$ & 1 & 2 & 1 & $-1/2$ & 0 \\ 
$(\nu_e,e)_R$ & 1 & 1 & 2 & $-1/2$ & 0 \\
\hline
$\Phi_L=(\phi_L^+,\phi_L^0)$ & 1 & 2 & 1 & 1/2 & 0 \\ 
$\Phi_R=(\phi_R^+,\phi_R^0)$ & 1 & 1 & 2 & 1/2 & 0 \\ 
\hline
$(N,E)_{L,R}$ & 1 & 2 & 1 & $-1/2$ & 1 \\ 
$(N',E')_{L,R}$ & 1 & 1 & 2 & $-1/2$ & 1 \\ 
$S_{L,R}$ & 1 & 1 & 1 & $0$ & 1 \\ 
\hline
$\chi^-$ & 1 & 1 & 1 & $-1$ & 1 \\
$\eta^-$ & 1 & 1 & 1 & $-1$ & 2 \\
$\sigma$ & 1 & 1 & 1 & 0 & 2 \\
\hline
\end{tabular}
\caption{Fermion and scalar content of left-right model with $U(1)_D$.}
\end{table}
The one-loop diagram for the electron mass is given in Fig.~1, 
\begin{figure}[htb]
\vspace*{-5cm}
\hspace*{-3cm}
\includegraphics[scale=1.0]{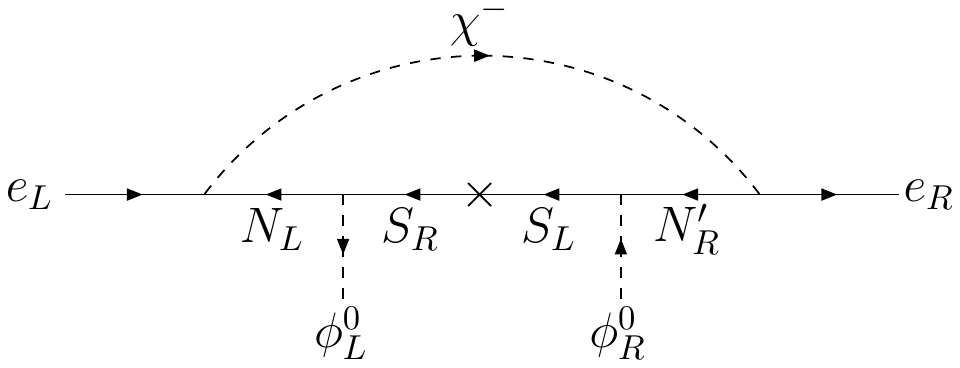}
\vspace*{-21.5cm}
\caption{Scotogenic electron mass.}
\end{figure}
with the 
understanding that there are two additional insertions linking $N_L$ to 
$N'_R$ as shown in Fig.~2.

\begin{figure}[htb]
\vspace*{-5cm}
\hspace*{-3cm}
\includegraphics[scale=1.0]{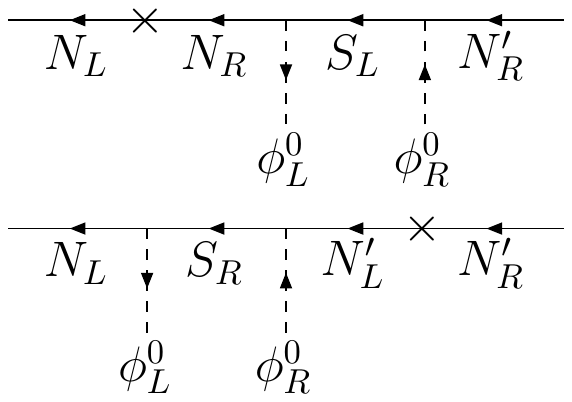}
\vspace*{-21.5cm}
\caption{Additional insertions to electron mass.}
\end{figure}
Assuming that $N'$ is the heaviest particle in the loop, the electron 
mass is then approximately given by
\begin{equation}
m_e \sim {f^4 v_L v_R \over 16 \pi^2 m_{N'}},
\end{equation}
where $f$ is a typical Yukawa coupling and $v_{L,R}/\sqrt{2}$ are the 
vacuum expectation values of $\phi^0_{L,R}$. 
Note that lepton number $L$ may be defined to be transmitted by $\chi^-$ 
across the loop.  Note also that neutrinos remain massless in one loop 
because the corresponding transmitter $\chi^0$ with $D=1$ is absent.  With 
the help of a second set of $(N,E)_{L,R}$, $S_{L,R}$, and $(N',E')_{L,R}$ 
fermions, as well as a charged scalar $\eta^-$ with $D=2$, two neutrinos 
will acquire three-loop Dirac masses as shown in Fig.~3, again with two 
additional insertions.
\begin{figure}[htb]
\vspace*{-5cm}
\hspace*{-3cm}
\includegraphics[scale=1.0]{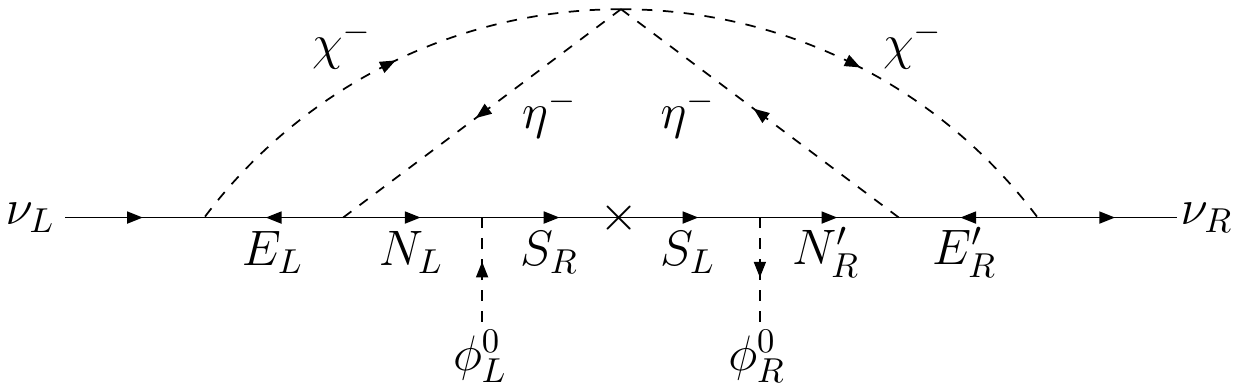}
\vspace*{-21.5cm}
\caption{Scotogenic Dirac neutrino mass.}
\end{figure}

A very rough estimate of the neutrino mass is then
\begin{equation}
m_\nu \sim {\lambda f^6 v_L v_R \over (16 \pi^2)^3 m_{N'}},
\end{equation}
where $\lambda$ is the quartic coupling of $(\chi^+ \chi^-)(\eta^+ \eta^-)$. 
Hence $m_\nu/m_e \sim \lambda f^2/(16 \pi^2)^2$ which is naturally of order 
$10^{-7}$ or so, in agreement with data.  Since there are two copies each 
of $(N,E)$, $S_{L,R}$, and $(N',E')$, two neutrinos will get mass.  The 
third neutrino gets a negligible mass from $W_{L,R}$ exchange, in analogy 
to the $2-W$ exchange~\cite{bm88} in the SM for Majorana neutrinos.  Other 
fermions may acquire seesaw Dirac masses~\cite{dw87} 
or radiatively~\cite{m20,m20-1}.

Whereas Fig.~3 represents the realization of the dimension-five operator 
\begin{equation}
{\cal L}^\nu_5 = {f^\nu_{ij} \over 2 \Lambda} (\nu_{iL} \phi_L^0 - l_{iL} 
\phi_L^+) (\bar{\nu}_{jR} \bar{\phi}_R^0 + \bar{l}_{jR} \phi_R^-), 
\end{equation}
Fig.~1 corresponds to
\begin{equation}
{\cal L}^e_5 = {f^e_{ij} \over 2 \Lambda} (\nu_{iL} {\phi}_L^- + 
{e}_{iL} \bar{\phi}_L^0) (\bar{\nu}_{jR} {\phi}_R^+ - \bar{e}_{jR} \phi_R^0). 
\end{equation}
These are the left-right analogs of Eq.~(1).  In previous applications, 
the electron mass often comes from a tree-level realization of ${\cal L}^e_5$, 
and only ${\cal L}^\nu_5$ is radiative in origin.  Recently~\cite{m20,m20-1}, 
both operators are derived in one loop, in which case there is no real 
understanding for why the Dirac neutrino masses are so much smaller than 
the electron mass.  Here the first example of one-loop electron mass and 
three-loop Dirac neutrino masses is presented.  Note also that Fig.~3 is 
a close analog to that~\cite{knt03} for a Majorana neutrino, known already 
many years ago.

\noindent \underline{\it Higgs and Gauge Sectors}~:~
The Higgs sector is identical to that of Ref.~\cite{m20-1}, consisting of 
scalars $\Phi_{L,R}$ and $\sigma$.  Although $\sigma$ now has $D=2$ instead 
of $D=3$, it does not affect the resulting Higgs potential, i.e.
\begin{eqnarray}
V &=& -\mu_L^2 \Phi_L^\dagger \Phi_L -\mu_R^2 \Phi_R^\dagger \Phi_R 
-\mu_\sigma^2 \sigma^* \sigma + {1 \over 2} \lambda_L (\Phi_L^\dagger 
\Phi_L)^2 + {1 \over 2} \lambda_R (\Phi_R^\dagger \Phi_R)^2 \nonumber \\ 
&+& {1 \over 2} \lambda_\sigma (\sigma^* \sigma)^2 + \lambda_{LR} 
(\Phi_L^\dagger \Phi_L)(\Phi_R^\dagger \Phi_R) + \lambda_{L\sigma}
(\Phi_L^\dagger \Phi_L)(\sigma^* \sigma) + \lambda_{R\sigma} 
(\Phi_R^\dagger \Phi_R)(\sigma^* \sigma).
\end{eqnarray}
After the spontaneous breaking of 
$SU(2)_L \times SU(2)_R \times U(1)_{B-L} \times U(1)_D$, the only physical 
scalars left are the real parts of $\phi^0_{L,R}$ and $\sigma$.  Let
\begin{equation}
\Phi_L = \pmatrix{0 \cr (v_L + h_L)/\sqrt{2}}, ~~~ \Phi_R = \pmatrix{0 \cr 
(v_R + h_R)/\sqrt{2}}, ~~~ \sigma = {1 \over \sqrt{2}} (v_D + h_D),
\end{equation}
then the $3 \times 3$ mass-squared matrix spanning $(h_L,h_R,h_D)$ is
\begin{equation}
{\cal M}^2_h = \pmatrix{\lambda_L v_L^2 & \lambda_{LR} v_L v_R & 
\lambda_{L\sigma} v_L v_D \cr \lambda_{LR} v_L v_R & \lambda_R v_R^2 & 
\lambda_{R\sigma} v_R v_D \cr \lambda_{L\sigma} v_L v_D & \lambda_{R\sigma} 
v_R v_D & \lambda_\sigma v_D^2}.
\end{equation}

In the gauge sector, the $Z_D$ boson gets a mass equal to $2 g_D v_D$. 
The charged $W^\pm_{L,R}$ masses are $g_L v_L$ and 
$g_R v_R$.  The $Z,Z'$ mass-squared matrix is 
\begin{equation}
{\cal M}^2_{Z,Z'} = e^2 \pmatrix{v_L^2/x(1-x) & v_L^2/(1-x)\sqrt{1-2x} \cr 
v_L^2/(1-x)\sqrt{1-2x} & (1-x)v_R^2/x(1-2x) + xv_L^2/(1-x)(1-2x)},
\end{equation}
where $e^{-2} = g_L^{-2} + g_R^{-2} + g_B^{-2}$, and $g_L=g_R$ with 
$x=\sin^2 \theta_W$.  The $Z-Z'$ mixing is then about 
$x\sqrt{1-2x} v_L^2/(1-x)^2 v_R^2$, which is constrained by  
the experimental bound of about $10^{-4}$~\cite{pdg18}. 

The dark $U(1)_D$ gauge symmetry is broken by two units through the 
complex singlet scalar $\sigma$, which couples to $S_L S_L$ and $S_R S_R$.  
This means that gauged $U(1)_D$ breaks to $Z_2$ dark parity $(-1)^D$, which 
is exactly conserved, allowing thus the lighter of the four Majorana fermions 
from the $(S_L,S_R)$ sector to become dark matter.

\noindent \underline{\it Dark Sector}~:~
The dark fermions $(N,E)$ and $(N',E')$ are assumed heavier than $S$.  Since 
$N$ and $N'$ mix with $S$ through $\phi^0_{L,R}$, they decay to $Sh_{L,R}$ 
and $E,E'$ decay to $S W^-_{L,R}$.  The dark scalar $\chi^-$ decays to 
$eN$ or $\nu E$, whereas $\eta^-$ decays to $E_1 N_2$ or $N_1 E_2$.  
The $2 \times 2$ mass matrix spanning $(\bar{S}_L,S_R)$ of the lighter 
of the two sets of $S_{L,R}$ is of the form
\begin{equation}
{\cal M}_S = \pmatrix{f_L v_D & m_S \cr m_S & f_R v_D}.
\end{equation}
Assuming $f_L=f_R=f$ and neglecting the $N-S$ and $N'-S$ mixings for now, 
this yields two Majorana fermions of masses $|m_S \pm fv_D|$, each coupling 
to $h_D$.  Let $S_1$ be the lighter, i.e. the dark-matter candidate, with 
mass $m_{S_1} = |m_S-fv_D|$.  Let $m_{h_D} < m_{S_1}$, then $S_1$ will 
annihilate to $h_D$, as shown in Fig.~4.  The first diagram is also 
accompanied by its $u-$channel counterpart, which has the same amplitude 
in the limit that $S_1$ is at rest.
\begin{figure}[htb]
\vspace*{-5cm}
\hspace*{-3cm}
\includegraphics[scale=1.0]{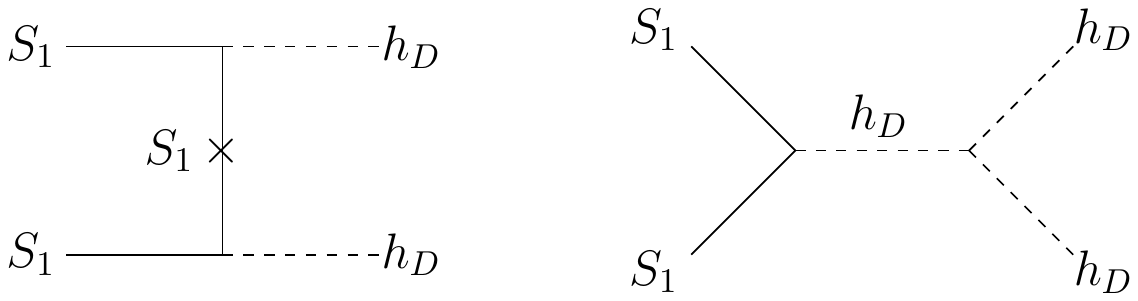}
\vspace*{-21.5cm}
\caption{$S_1 S_1$ annihilation to $h_D h_D$.}
\end{figure}
Let $x = m_{h_D}/m_{S_1}$ and using $m^2_{h_D} = \lambda_\sigma v_D^2$, 
this cross section at rest multiplied by relative velocity is
\begin{equation}
\sigma_{ann} \times v_{rel} = {\sqrt{1-x^2} \over 128 \pi} 
\left| {2f^2 \over m_{S_1} (1+x^2)} - {3fx^2 \over v_D(4-x^2)} \right|^2.
\end{equation}
As an example, let $m_{S_1}=v_D=1$ TeV, $m_{h_D} = 500$ GeV, then the 
canonical value $3 \times 10^{-26}~{\rm cm}^3/{\rm s}$ for the correct 
relic abundance of dark matter in the Universe is obtained if $f = 0.89$.  
This in turn implies $m_S = 1.89$ TeV.

As for direct detection, $S_1$ couples to the SM Higgs $h_L$ through mixing 
with $h_D$ which is $\lambda_{L\sigma} v_L v_D/m^2_{h_D}$.  For $m_{S_1} = 1$ 
TeV, the spin-independent cross section of dark matter scattering off a 
xenon nucleus is bounded by~\cite{xenon18} $10^{-45}$~cm$^2$.  This puts a 
limit of $4 \times 10^{-4}$ on $\lambda_{L\sigma}$ for $v_D=1$ TeV and 
$m_{h_D} = 500$ GeV.  With this value of $\lambda_{L\sigma}$, the 
decay $h_D \to h_L h_L$ has a lifetime of $2.4 \times 10^{-19}$~s, 
certainly short enough to allow it to be in thermal equilibrium with 
the SM particles.  Since $S_1$ mixes with $N$ through $\phi_L^0$ and $N'$ 
through $\phi_R^0$, it couples to the vector gauge bosons $Z$ and $Z'$, which 
interact with quarks.  However, $S_1$ is a Majorana fermion, so this 
cross section is zero in the limit that it is at rest.

\noindent \underline{\it Conclusion}~:~
In the conventional left-right gauge extension of the standard model with 
just one $SU(2)_L$ scalar doublet and one $SU(2)_R$ scalar doublet, it is 
shown how a radiative electron mass arises in one loop, as well as related 
Dirac neutrino masses in three loops.  The particles in the loop belong 
to a dark sector with $U(1)_D$ gauge symmetry.  The spontaneous breaking 
of $U(1)_D$ leads to a viable Majorana fermion dark-matter candidate.

\noindent \underline{\it Acknowledgement}~:~
This work was supported in part by the U.~S.~Department of Energy Grant 
No. DE-SC0008541.

\bibliographystyle{unsrt}

\begin{thebibliography}{99}
\bibitem{w79} S. Weinberg, Phys. Rev. Lett. {\bf 43}, 1566 (1979).
\bibitem{m98} E. Ma, Phys. Rev. Lett. {\bf 81}, 1171 (1998).
\bibitem{bh18} G. Bertone and D. Hooper, Rev. Mod. Phys. {\bf 90}, 045002 
(2018).
\bibitem{m06} E. Ma, Phys. Rev. {\bf D73}, 077301 (2006).
\bibitem{mp17} E. Ma and O. Popov, Phys. Lett. {\bf B764}, 142 (2017).
\bibitem{bms03} B. Brahmachari, E. Ma, and U. Sarkar, Phys. Rev. Lett. 
{\bf 91}, 011801 (2003).
\bibitem{dw87} A. Davidson and K. C. Wali, Phys. Rev. Lett. {\bf 59}, 393 
(1987).
\bibitem{m88} R. N. Mohapatra, Phys. Lett. {\bf B201}, 517 (1988).
\bibitem{m89} E. Ma, Phys. Rev. Lett. {\bf 63}, 1042 (1989).
\bibitem{ms18} E. Ma and U. Sarkar, Phys. Lett. {\bf B776}, 54 (2018).
\bibitem{m20} E. Ma, Phys. Lett. {\bf B811}, 135971 (2020). 
\bibitem{m20-1} E. Ma, arXiv:2012.03128 [hep-ph].
\bibitem{bm88} K. S. Babu and E. Ma, Phys. Rev. Lett. {\bf 61}, 674 (1988).
\bibitem{knt03} L. M. Krauss, S. Nasri, and M. Trodden, Phys. Rev. {\bf D67}, 
085002 (2003). 
\bibitem{pdg18} M. Tanabashi {\it et al.} (Particle Data Group), Phys. Rev. 
{\bf D98}, 030001 (2018).
\bibitem{xenon18} E. Aprile {\it et al.} (XENON Collaboration), Phys. Rev. 
Lett. {\bf 121}, 111302 (2018).



\end{thebibliography}

\end{document}